# Chemically consistent evolution of galaxies:

## Spiral galaxy models compared to DLA systems

**Ulrich Lindner, Uta Fritze – von Alvensleben, Klaus J. Fricke**

Universitätssternwarte, Geismarlandstraße 11, D–37083 Göttingen, Germany
e-mail: ulindner@uni-sw.gwdg.de

**Abstract.** We have extended our chemical and cosmological galaxy evolution model to calculate the abundance evolution for altogether 16 different elements in spiral galaxies in a chemically consistent way which is a considerable step towards a more realistic galaxy modeling. All observed element abundances in DLA systems have been compiled. The comparison with our model calculations yields the following results.

Together with the fact that our models well reproduce observed average HII region abundances in all spiral types the conformity between observed and calculated abundances over the redshift range from $z \sim 4.5$ through $z \sim 0.4$ indicates that DLA galaxies may well evolve into the full range of present–day spiral galaxies from Sa through Sd.

Comparison of our chemically consistent models with models using only solar metallicity input physics shows that differences in the redshift evolution are small for some elements but large for others. For those elements with large differences the chemically consistent models provide significant better agreement with observed DLA abundances.

For typical spiral galaxies the star formation histories of our models clearly bridge the gap between high redshift DLA systems and the nearby spiral galaxy population. The slow redshift evolution of DLA abundances is understood in terms of the long star formation timescales in galactic and proto-galactic disks. The large scatter of observed abundances in DLAs of similar redshift is rather explained by the range of SFRs among early and late type spirals.

Towards lower redshift $z \leq 1.5$ our models indicate that early type spirals drop out of the DLA samples as their gas content falls below $\sim 50\,\%$. Implications for optical identification are discussed.

## 1. Introduction

Abundances of various heavy elements in Damped Ly$\alpha$ (**DLA**) absorbers are being determined since many years now, first mainly in the redshift range $z \sim 2\ldots 3$ where the DLA lines are accessible from the ground. Abundances have been determined from curve of growth analysis e.g. of the low ionisation lines of ZnII, CrII, FeII, SiII, and many others, which are found associated with the DLA line. The HST key project Quasar Absorption Lines (Bahcall *et al.* 1996, 1993) extended the possible range of DLA detections towards lower redshifts. In particular three observational facts still keep challenging our understanding of the nature of DLA systems.

– The number of low redshift DLA systems detected is much smaller than expected.
– The redshift evolution of heavy element abundances in DLA systems is very weak, in particular if compared to the strong redshift evolution of the narrow high ionisation CIV QSO absorption systems.
– The scatter of abundances observed among various DLA systems at any given redshift is very large.

In 1995 we presented a first comparison of our chemical and cosmological galaxy evolution models for spiral galaxies of various types with DLA abundances (Fritze - v. Alvensleben & Fricke 1995b, see also Fritze - v. Alvensleben 1995a). At that time, however, observational abundance determinations were not very precise yet, giving lower and upper limits only in many cases and we felt somewhat uncomfortable using stellar yields calculated for solar metallicity stars to compare with high redshift DLAs which typically have low metallicities $\frac{1}{100} \cdot Z_\odot \ldots Z_\odot$.

Meanwhile, the situation has improved considerably. KECK HIRES, HST GHRS, and WHT spectra give completely resolved absorption line profiles for a large number of lines in many DLAs. In these cases the apparent optical depth method, which does not require any assumption about (the functional form of) the velocity distribution in the absorbing gas, allows for precise abundance determinations (e.g. Pettini *et al.* 1994, Prochaska & Wolfe 1996, Lu *et al.* 1996). This method works well even for kinematically complex multi-component profiles. It also allows to correct for saturated lines, in which case still, however, only lower limits for the respective element abundances can be obtained.

On the theoretical side, stellar yields for a set of different metallicities from $Z = 0$ to $Z_\odot$ have become available (Woosley & Weaver 1995, van den Hoek & Groenewegen 1997, Marigo *et al.* 1998, Portinari *et al.* 1998) which we now use in our modeling. We have developed a method to model the chemical evolution of ISM abundances in galaxies in a chemically consistent way, very much in parallel to the chemically consistent treatment of the photometric (Einsel *et al.* 1995, Fritze - v. Alvensleben *et al.* 1996, Möller *et al.* 1997) and spectral (Möller *et al.* 1998) evolution. For a galaxy, which always is a composite system in terms of stellar metallicities and ages we follow the chemical evolution of successive generations of stars using yields and stellar lifetimes appropriate for their respective initial metallicities.

We use a set of star formation histories (**SFH**s) appropriate for spiral galaxies of various types that provide a successful description not only of the detailed spectral properties of the respective nearby template galaxies in the optical, their average colors from U through K, their emission line properties but also of their redshift evolution back to $z \sim 1$ as far as accessible via type-dependent redshift surveys (cf. Möller *et al.* 1996 and 1998). We show that with these SFHs the chemically consistent chemical and cosmological evolution models give good agreement of the model abundances at $z = 0$ with observed HII region abundances of nearby galaxies. We thus expect our models to also be able to describe the redshift evolution of the ISM abundances in spiral or proto-spiral galaxies and thus to provide a tool to bridge the gap between DLA absorbers at high redshift and the local galaxy population.

We present our models, their basic parameters, and the input physics we use in section 2. All available DLA abundance data are compiled and described in section 3 where we also discuss their various degrees of reliability. In section 4, we present a detailed comparison of the redshift evolution of various element abundances (Fe, Si, Zn, Cr, Ni, S, Al, Mn) as given by our models for various spiral types with all the available data for DLAs. Results are discussed as to our understanding of the weak redshift evolution of observed DLA abundances, the scatter they show at any given redshift, the question as to the nature of the absorber galaxy/protogalaxy population as well as to the importance of the chemically consistent approach. In section 5 we use the comparison between models and observations to discuss the properties of the DLA absorbing galaxy population and their redshift evolution, derive some implications and present predictions for optical identifications of DLA galaxies. We summarize our main results in section 6.

## 2. Chemically consistent galaxy evolution models

Our galaxy evolution model has been described in detail earlier by Fritze – von Alvensleben 1989, Krüger *et al.* 1991, Fritze – von Alvensleben & Gerhard 1994 and Lindner *et al.* 1996. In the following brief outline we concentrate on the chemical evolution, especially on our new concept of chemical consistency which is a considerable step towards a more realistic galaxy modeling. Chemically consistent models account for the increasing initial stellar metallicities of successive generations of stars and use several sets of stellar evolutionary tracks, stellar lifetimes and yields, color calibrations and spectra appropriate for the different metallicity subpopulations present in any type of galaxy. While single burst stellar populations, like star clusters, are well described by one common (age and) metallicity for all stars, the stars in any system with an extended or more complex star formation history (**SFH**) have a dispersion not only in age but also in initial metallicity.

Our chemically consistent models describe the first stars forming in a (proto–) galaxy using the lowest metallicity stellar tracks, lifetimes, yields, etc. and consistently use input data bases for higher metallicity as the ISM abundance increases. Our models, however, do not include any dynamical aspects, we have to assume that the gas is always well mixed (cf. section 2.2).

Chemically consistent models can now be developed because sufficiently complete and homogeneous sets of physical input data are becoming available for a range of metallicities, including metallicity dependent stellar yields. With our new model approach and extensive input data bases we calculate in detail the time and redshift evolution of abundances for a large number of different elements including SNI contributions from carbon deflagration white dwarf binaries.

Chemically consistent models similar in principle to the ones presented here were used by Timmes *et al.* (1995) using the same Woosley & Weaver (1995) yields for massive stars but older yields (Renzini & Voli (1981)) for stars with $m < 8\,M_\odot$. Recent chemically consistent models by Portinari *et al.* (1998) use the Padova set of stellar input physics. Both approaches aim at describing the *Galactic* enrichment history by comparing to observed *stellar* abundance patterns.

### 2.1. General description of models

Starting from an initial gas cloud of mass $G(t = 0) = M_{\rm tot}$ stars are formed continuously in a 1–zone model according to a given star formation law. The distribution of the total astrated mass to discrete stellar masses in the range 0.15 $M_\odot$ ... 40 $M_\odot$ is described by a Scalo 1986 initial mass function (IMF) as specified in equation (1).

$$\phi(m) = \begin{cases} \phi_1 * m^{-(1+x_1)} & m_{down} \le m \le m_{12} \\ \phi_2 * m^{-(1+x_2)} & m_{12} < m \le m_{23} \\ \phi_3 * m^{-(1+x_3)} & m_{23} < m \le m_{up} \end{cases} \quad (1)$$

with normalization factors $\phi_1 = \phi_2 = 0.1974$ and $\phi_3 = 0.3581$, slopes $x_1 = 0.25$, $x_2 = 1.35$, and $x_3 = 2.0$ and mass limits $m_{down} = 0.15$, $m_{12} = 1.0$, $m_{23} = 2.5$, and $m_{up} = 40.0\,M_\odot$. The IMF is normalized to $\int_{m_{down}}^{m_{up}} \Phi(m) \cdot m \cdot dm = 0.5$ to account for a fraction of astrated material stored away from enrichment and recycling processes in substellar objects $m < m_{down} = 0.15 M_\odot$ (Bahcall et al. 1985). This normalization at the same time brings our model M/L values after a Hubble time into good agreement with observed M/L values for the respective galaxy types. The influence on our results from the use of this Scalo IMF as compared to a single Salpeter slope $x = 1.35$ is discussed in Section 2.3.5.

For galaxies of various spectral types, we use different parametrisations of their star formation rates (SFR) following Sandage's (1986) semi-empirical determinations. SFRs $\Psi(t)$ in spiral galaxies are assumed to be linear functions of the gas content $g(t) := G(t)/(M_{tot} [10^9\, M_\odot])$ with characteristic time scales $t_*$ ranging from about 2 to 10 Gyr for Sa, Sb, and Sc, respectively, and a constant rate for Sd galaxies (cf. equation (2)).

$$\Psi(t) = \begin{cases} g(t) * 0.4 & , t_* = 2\,\text{Gyr} & \text{Sa} \\ g(t) * 0.3 & , t_* = 3\,\text{Gyr} & \text{Sb} \\ g(t) * 0.1 & , t_* = 10\,\text{Gyr} & \text{Sc} \\ M_{tot}\,[10^9\,M_\odot] * 3.5\,10^{-2}, & t_* = 16\,\text{Gyr} & \text{Sd} \end{cases} \quad (2)$$

The total mass is assumed to be constant ($M_{tot}$ = const), since we restrict ourselves to closed box models. All masses are given in solar units ($M_\odot$). The characteristic timescale for SF $t_*$ is defined by $\int_0^{t_*} \Psi \cdot dt = 0.63 \cdot G(t = 0)$. Gas recycling due to stellar winds, supernovae and planetary nebula is included consistently accounting for the finite stellar lifetimes $\tau_m$ of stars of mass m, i. e., no Instantaneous Recycling Approximation is used (cf. section 2.2).

Since dynamical effects are not included in our models, we cannot account for the internal structure or gradients in spiral galaxies or DLA absorbers. Our closed box models do not allow for galactic winds which clearly are important for dwarf galaxies but presumably not for spiral galaxies or their massive DLA progenitors (cf. A. Wolfe (1995), Wolfe & Prochaska (1997), and references therein).

For these SFRs our chemically consistent spectrophotometric models (cf. Möller et al. 1996, Fritze – v. Alvensleben et al. 1996) give detailed agreement, not only with average broad band colors observed for the respective galaxy types from U through K, but also with detailed emission and absorption features of the template spectra from Kennicutt's (1992) atlas. They also give agreement with the observed redshift evolution of galaxy colors for the respective types at least up to $z \sim 1$, i.e. over roughly a third of the Hubble time and with observations of optically identified QSO absorbers (Lindner et al. 1996).

Although our simple 1-zone models are not able to account for any abundance gradients along (proto-)galactic disks, the range of observed average HII region abundances $-0.23 \le [O/H] \le +0.17$ for nearby spiral types Sa through Sd as given by Zaritsky et al. (1994) and Oey & Kennicutt (1993) is well covered by our Sa through Sd models at $z = 0$ (cf. Fig. 6 in section 2.3.5). Of particular interest for the comparison with DLA abundances which – most probably – probe the outer regions of the absorbing galaxies are the HII region abundances measured by Ferguson et al. (1998). For three late type spirals with large HI-to-optical sizes they find $[O/H] \sim 10 - 15\%$ solar at 1.5 – 2 optical radii. For their galaxies this corresponds to off-center distances between 12 and 42 kpc where the HI column density still is several $10^{20}$ cm$^{-2}$. Impact parameter found for optically identified DLA absorber galaxies typically are in the range 10 – 20 kpc.

*2.2. Input physics*

Besides the two basic parameters of our models, IMF and SFR, **stellar yields** $p_i(m)$ for elements i, **stellar remnant masses** $m_{rem}$, and **stellar lifetimes** $\tau_m(m)$ are required as input physics for our models. These data are needed to calculate the total mass E(t) ejected by stars above the turn-off mass $m_t$

$$E(t) = \int_{m_t}^{m_{up}} (m - m_{rem}) \cdot \Phi(m) \cdot \Psi(t - \tau_m(m)) \cdot dm \quad (3a)$$

that, together with the SFR, determines the time evolution of the gas mass in our closed box models:

$$\frac{dG}{dt} = -\Psi(t) + E(t) \quad (3b).$$

Our models only aim at describing average gas phase abundances without accounting for the multiphase nature of the ISM, they assume perfect and instantaneous mixing of the material rejected by the stars. For a simple 1-zone model, the abundance $X_i$ for each element i is calculated from equation (3c)

$$\frac{dX_i}{dt} = (E_i(t) - E(t) \cdot X_i(t))/G(t) \quad (3c)$$

with

$$E_i = E_i^1 + E_i^2 + E_i^3 + E_i^4$$

.

We follow the formalism outlined by Matteucci & Greggio (1986) and Matteucci & Tornambè (1987) to split up the IMF in the mass range between 3 and 16 $M_\odot$ into some fraction A of binary stars that give rise to SNIa in the carbon deflagration white dwarf binary scenario and a fraction $(1-A)$ of single stars. The term

$$E_i^1(t) = \int_{m_t}^{1.5 M_\odot} m \cdot p_i(m) \cdot \Psi(t-\tau_m(m)) \cdot \Phi(m) \cdot dm$$

describes the ejection contribution to an element i from single low mass stars,

$$E_i^2(t) = A \cdot \int_{3M_\odot}^{16M_\odot} \int_{\mu_m}^{0.5} f(\mu) \cdot m_p \cdot p_i^{SNIa}(m_p)$$
$$\cdot \Psi(t-\tau_m(m_s)) \cdot d\mu \cdot \Phi(M_B) \cdot dM_B$$

the respective contribution of type Ia SNe in binary systems with binary mass $M_B = m_p + m_s$. After its lifetime the primary star $m_p$ is assumed to transform into a white dwarf, ejecting its envelope into the ISM. Later, the secondary $M_s$ evolves into a red giant, fills its Roche lobe and all the material is assumed to flow onto the primary which eventually reaches the Chandrasekhar limit and gives rise to a carbon deflagration SNIa event. The parameter A gives the fraction of stars in the mass range 1.5 – 8 $M_\odot$ that finally give rise to a SNIa event, $\mu := \frac{m_s}{M_B}$ and $\mu_m := \max\{\frac{m_s(t)}{M_B}, \frac{M_B - 8M_\odot}{M_B}\}$ is – in analogy to the turn-off mass for single stars – the smallest mass fraction that contributes to SNIa at time t. $f(\mu) \sim \mu^2$ is an assumed distribution function for binary relative masses. As proposed by Matteucci & Greggio, the parameter A is fixed by the requirement that for a Milky Way model Sbc the resulting SNIa rate at $\sim 12$ Gyr is equal to the observed one (Cappellaro *et al.* 1997). This gives us A = 0.1 which we use for all galaxy types. The enrichment contribution of single stars in this mass range is given by

$$E_i^3(t) = (1-A) \cdot \int_{1.5M_\odot}^{8M_\odot} m \cdot p_i(m) \cdot \Psi(t-\tau_m(m)) \cdot \Phi(m) \cdot dm$$

and the SNII contributions from stars above 8 $M_\odot$ are described by

$$E_i^4(t) = \int_{8M_\odot}^{m_{up}} m \cdot p_i(m) \cdot \Psi(t-\tau_m(m)) \cdot \Phi(m) \cdot dm.$$

### 2.2.1. Cosmological model

To compare the results from our galaxy evolution calculations with the observed element abundances in DLA systems redshift dependent values are needed. To convert any evolution in time to a redshift evolution we adopt a Friedmann–Lemaître model with vanishing cosmological constant ($\Lambda_0 = 0$) and cosmological parameters $H_0 = 50$ km s$^{-1}$ Mpc$^{-1}$ and $\Omega_0 = 1$. As redshift of galaxy formation we chose $z_{form} = 5$. These parameters are in conformity with our spectrophotometric models (cf. Möller *et al.* 1996, Fritze – v. Alvensleben *et al.* 1996) and are used throughout the paper. The relation between redshift and time is then calculated directly obtained via (equation (4)) for the *Hubble–time* $T_H(z, H_0, \Omega_0)$.

$$t_{gal}(z) = T_H(z) - T_H(z_{form}) \quad (4)$$

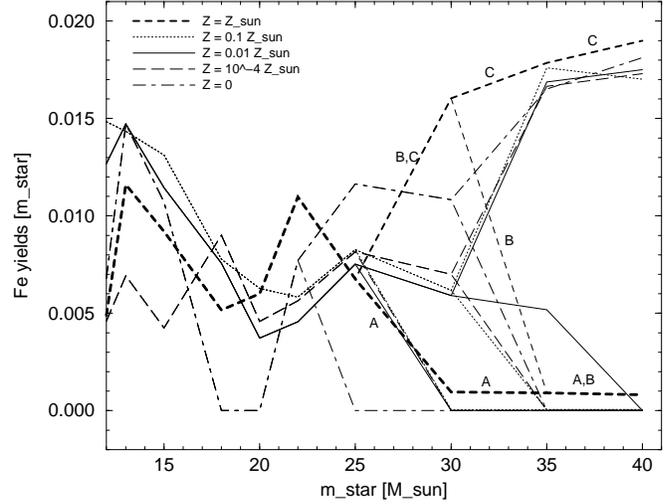

**Fig. 1.** Iron yields are given in units of stellar mass m_star as a function of stellar mass m_star (given in units of solar mass M_sun).

### 2.2.2. Supernova yields

Supernova explosions of type II (SN II), i.e. stars heavier than about ten solar masses, are the most productive suppliers of heavy elements to the interstellar medium (ISM). SN type I explosions also supply a considerable contribution to the ISM metallicity (see Nomoto *et al.* 1997) in case of some elements, i.e. Fe, Ni, Cr and Mn (cf. last row of Table 5); whereas single stars of intermediate and small mass ($m < 10\ M_\odot$) supply considerable contributions only to the abundances of elements C, N and O (cf. Table 6).

Woosley & Weaver (1995) have calculated nucleosynthetic yields of about 144 isotopes from altogether 32 elements (H, He, Li, Be, B, C, N, O, F, Ne, Na, Mg, Al, Si, P, S, Cl, Ar, K, Ca, Sc, Ti, V, Cr, Mn, Fe, Co, Ni, Cu, Zn, Ga, Ge) ejected from SN II explosions of progenitor stars with 12, 13, 15, 18, 20, 22, 25, 30, 35 and 40 solar masses. They considered five different initial metallicities ($Z = Z_\odot$, $Z = 0.1\ Z_\odot$, $Z = 0.01\ Z_\odot$, $Z = 10^{-4}\ Z_\odot$ and $Z = 0$) and three models with different explosion energies (labeled A, B and C) for very massive stars. For our models we use their SNII yields for 16 of the most abundant elements (H, He, C, N, O, Mg, Al, Si, S, Ar, Ca, Cr, Mn, Fe, Ni and Zn). Ejecta (given in solar mass units $M_\odot$) are listed in **Table 5** for all five initial metallicities summed over all isotopes for each element. In the last column of this table we see that differences $\Delta$ between the total ejection of all elements ($E_{tot}$) and the sum of ejecta from our 16 selected elements is negligible (between 0.3 % and 4.1 %).

Ejected masses of all isotopes of one element are added because we are only interested in total element abundances. Woosley & Weaver (1995) did not take into consideration any radioactive decay of isotopes after their pro-duction in the SN II explosion. In **Table 1** we list those radioactive decays of isotopes which considerably contribute to the abundances of Fe, Cr and Mn. All contributions of other isotopes are negligible for our purpose.

**Table 1.** Radioactive decays relevant for our yields

| | | |
|---|---|---|
| $^{56}$Ni $\Longrightarrow$ $^{56}$Fe | $^{53}$Fe $\Longrightarrow$ $^{53}$Cr | $^{55}$Co $\Longrightarrow$ $^{55}$Mn |
| $^{52}$Fe $\Longrightarrow$ $^{52}$Cr | $^{53}$Mn $\Longrightarrow$ $^{53}$Cr | $^{55}$Fe $\Longrightarrow$ $^{55}$Mn |

Rows containing results from model B and C are marked with $^*$ and $^{**}$ attached to the stellar mass in the first column of Table 5. We see that for larger explosion energy the ejected mass of heavy elements increases. ".0000" in Table 5 indicate that respective values are smaller than $1.0 \cdot 10^{-4}$.

For an overview of the influence of different initial metallicity we list in Table 5 as an example the SN II yields for a "typical" (i.e. 25 $M_\odot$) star (model A). Ejecta for Z = 0 are clearly smaller than for other metallicities. On the whole, Z=0 yields differ drastically from those for Z > 0 but these exceptional data do not affect our results because after the first few timesteps our program switches to higher metallicity input data. However, yields for other metallicities likewise do not show any general trend, neither with respect to initial stellar mass for fixed metallicity nor with respect to initial metallicity for fixed stellar mass, and this applies to all elements considered. In case of iron and carbon this is illustrated in Fig. 1 and Fig. 2.

In **Fig. 1** we present stellar iron yields for the five initial metallicities calculated by Woosley & Weaver 1995. Yields are given as a fraction of the total stellar mass $m_{star}$ and stellar masses are given in solar units $M_{sun}$. Lines do split at $m_* = 30$ and $35 M_\odot$ indicating different explosion models. In case of solar metallicity (Z = $Z_\odot$) the separate lines for model A, B and C are indicated for stellar mass larger than 25 $M_\odot$. We see that generally more mass is ejected for larger explosion energies. However, no distinct trend neither with increasing stellar mass nor with increasing metallicity can be found.

We use element yields from SNIa calculated from Nomoto's deflagration model W7 (Nomoto *et al.* 1997), which are presented in the last row of Table 5. These SNIa yields are available for solar metallicity only. However, no important metallicity dependence is expected for SNIa yields.

### 2.2.3. Yields from intermediate and low mass stars

Intermediate mass stars ($0.9 M_\odot < m < 8 M_\odot$) contribute little to the total metal enrichment of the ISM, they are, however, important for elements C, N and O. Stars of mass less than $0.9 M_\odot$ do not contribute any metals to the ISM. We use up-to-date stellar yields for three different initial metallicities (Z = $Z_\odot$, Z = $0.2 Z_\odot$ and Z = $0.05 Z_\odot$) calculated by van den Hoek & Groenewegen 1997. The data are listed in **Table 6** for stellar masses ranging from 0.9 $M_\odot$ to 8 $M_\odot$. Negative values indicate consumption instead of ejection of mass and "0.000" indicate values smaller than $1.0 \cdot 10^{-3}$. Yields of the same element but for different metallicities are arranged in neighboring columns of the table to make comparison easy. In case of elements C, N and O differences are quite significant but no trend can be found.

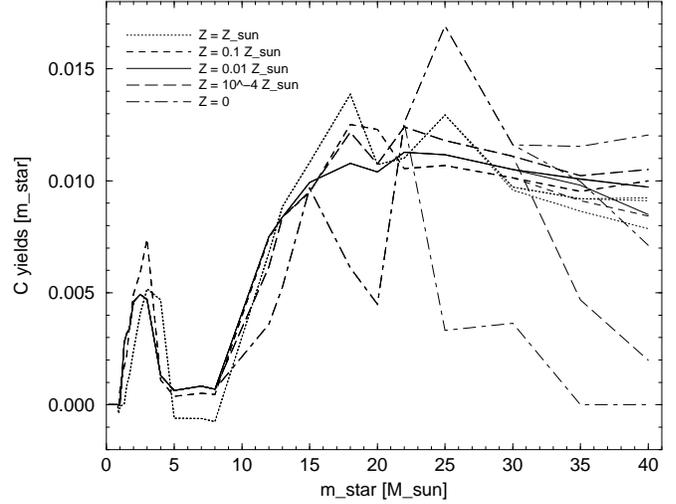

**Fig. 2.** Carbon yields (given in units of stellar mass m_star) as a function of stellar mass m_star (given in units of solar mass M_sun).

In **Fig. 2** we present stellar carbon yields in units of stellar mass $m_{star}$ for five initial metallicities calculated by Woosley & Weaver (1995). For stellar masses less than 8 $M_\odot$ the plot contains the results calculated by van den Hoek & Groenewegen (1997). Their initial metallicities differ from those used by Woosley & Weaver and we combine those metallicities which are next to each other as listed in **Table 2**.

**Table 2.** Metallicities combined from Woosley & Weaver (1995) for high mass stars and van den Hoek & Groenewegen (1997) for intermediate mass stars

| | | | | | |
|---|---|---|---|---|---|
| Woosley & Weaver | $Z_\odot$ | $0.1 Z_\odot$ | $0.01 Z_\odot$ | $10^{-4} Z_\odot$ | $0 Z_\odot$ |
| v.d. Hoek & G. | $Z_\odot$ | $0.2 Z_\odot$ | $0.05 Z_\odot$ | $0.05 Z_\odot$ | $0.05 Z_\odot$ |

As for iron in Fig. 1 carbon lines do split at $m_* = 25$ and $30 M_\odot$ indicating different explosion models and clearly more carbon is ejected for larger explosion energy. Likewise no distinct trend with increasing stellar mass or increasing metallicity is visible. These findings are typical not only in case of iron and carbon but for almost all elements under investigation.

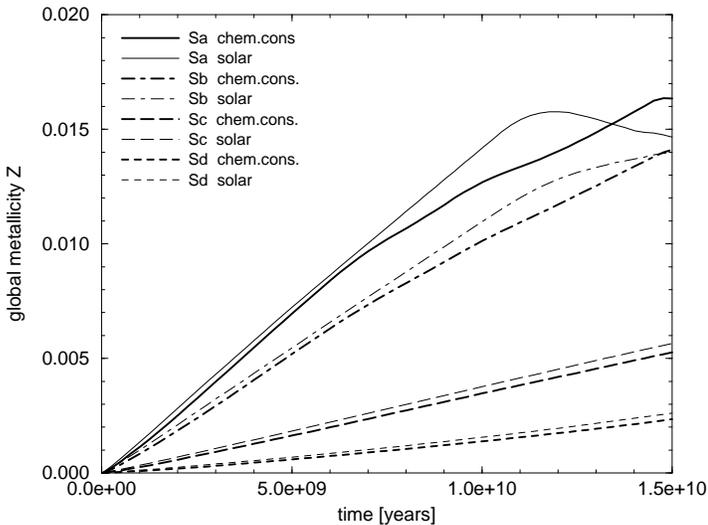

**Fig. 3.** Time evolution of global metallicity for Sa, Sb, Sc and Sd galaxies: comparison of results calculated with **chemical consistent** models vs. models using exclusively **solar** values.

### 2.2.4. Stellar remnants and lifetimes

Neither Woosley & Weaver (1995) nor van den Hoek & Groenewegen (1997) report any stellar lifetimes $\tau_m(m)$. Hence we adopt them from the stellar evolutionary tracks calculated by the Geneva group who gives lifetimes for two different initial metallicities $Z = Z_\odot$ and $Z = 0.05 \, Z_\odot$, listed in **Table 7**. For stellar masses less than 0.9 $M_\odot$ there is no difference for lifetimes between the different metallicities and lifetimes are equal to or larger than the Hubble time, anyway. Hence, only data for $m \geq 0.9 \, M_\odot$ are of interest. To coordinate with yields for different initial metallicities we use $\tau_1$ in case of $Z = Z_\odot$ and $\tau_2$ for all other metallicities.

Masses of stellar remnants have been calculated by Woosley & Weaver (1995) for massive stars and by van den Hoek & Groenewegen (1997) for intermediate mass stars. Their results are reported in **Table 8** and **Table ??**, respectively.

### 2.3. Discussion of models

Now the influence of different initial metallicities and explosion energies (described in the previous section 2.2) on the results of our chemical evolution models will briefly be discussed.

### 2.3.1. Chemically consistent models

During the evolution of any galaxy the ISM is continuously enriched with metals. Hence it is reasonable to assume that stars which are formed in early phases are very poor in metals and consequently we need to use input data (yields, remnants and lifetimes) of very low metallicity in the beginning of the galaxy evolution. With increasing time the metal content of the ISM is growing and after each time step the actual metallicity is determined to select the appropriate input data. In earlier evolution models only solar metallicity data have been available. Chemically consistent models take into account increasing metal enrichment of the ISM from which successive generations of stars are born and hence are more realistic than models using solar data. A comparison of chemically consistent calculations only with results from models using solar metallicity exclusively for Sa, Sb, Sc and Sd galaxies is shown in **Fig. 3**. We see that chemically consistent models in general produce less metals than calculations with solar input data because metallicity dependent stellar yields are smaller than their solar counterparts. In the following we will omit curves of Sb and Sc galaxies because they lie between those of Sa and Sd galaxies.

The average ISM metallicity of our Sb model after a Hubble time is seen to be about $2/3 \, Z_\odot$ in good agreement with recent ISM abundance determinations (cf. Cardelli & Meyer 1997, Sofia *et al.* 1997, and references therein), H II region abundances (e.g. Vilchez & Esteban 1996) for the solar neighborhood and with B–star abundances (e.g. Kilian *et al.* 1994, Kilian–Montenbruch *et al.* 1994).

### 2.3.2. The influence of stellar yields

As discussed in section 2.2.2 Woosley & Weaver's yields show no clear trends neither with stellar mass at fixed metallicity Z nor with Z at fixed stellar mass. In particular yields for $Z = 0$ differ drastically from those with $Z \neq 0$.

We decided to use Woosley & Weaver's data because they give yields for five different metallicities which is important for our concept of chemical consistent models. It should be mentioned that there are yield data from other authors.

Thielemann *et al.* (1996) published SN II yields for solar initial metallicity which to some extent differ from those of Woosley & Weaver. For a detailed investigation of the effects of these differences we refer the reader to D. Thomas *et al.* (1998). Portinari *et al.* (1998) take mass loss by stellar winds into account and give stellar yields for a few elements for five initial metallicities.

The impact of yield uncertainties on our results is hard to quantify. Even significant changes for a star of given mass and metallicity, however, do hardly affect the global evolution due to the smoothing power of the IMF.

From a comparison of the stellar yields given by various authors we conclude that while yield differences may have strong impact on abundance ratios of certain elements – which we do not attempt to interpret – they will not strongly affect the abundance evolution and hence our conclusions.

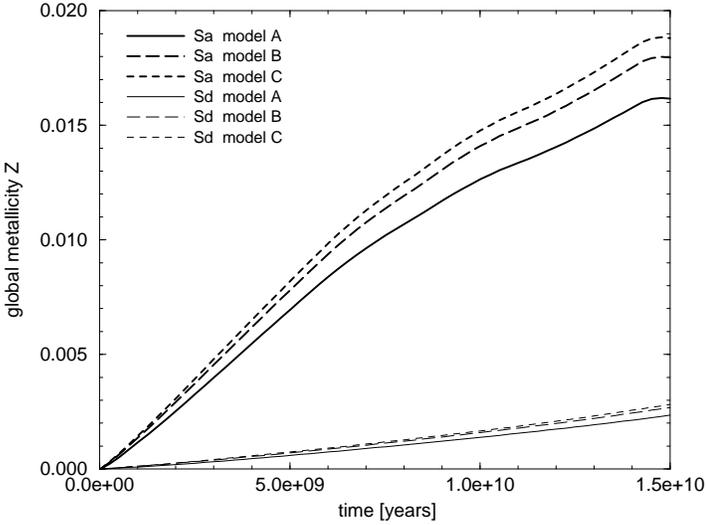 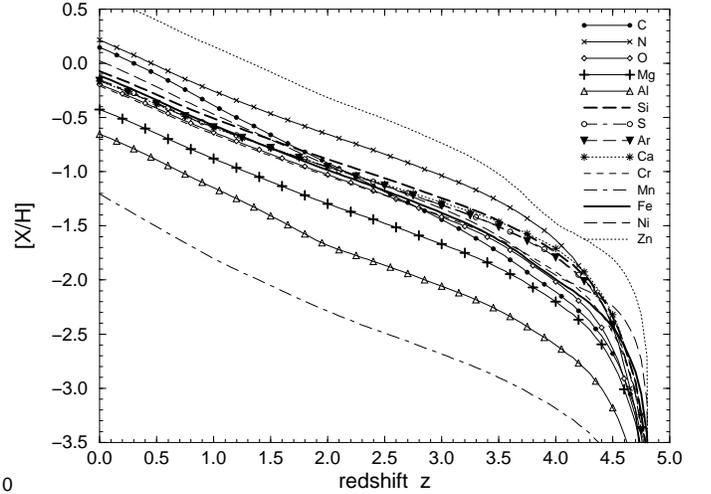

**Fig. 4.** Time evolution of global metallicity for Sa and Sd galaxies: comparison of results calculated with chemical consistent models using different energies for SN II explosions: models A, B and C from Woosley & Weaver (1995)

**Fig. 5.** redshift evolution for various element abundances [X/H] in Sa galaxies. Elements X are listed in the legend.

Yield uncertainties may slightly change our enrichment calculations for some elements, e.g. for Fe (or for typical wind elements C, N, O, which, however, we do not discuss since there are very few precise DLA data), but certainly not to the extent as to affect our conclusions which are based on a series of elements for many of which stellar yields are not controversial.

2.3.3. Different explosion energies for SN II

A comparison of results from chemically consistent evolution models for Sa and Sd galaxies using different explosion energies for SN II yields calculated by Woosley & Weaver (1995) is shown in **Fig. 4**. We see that the curves representing the time evolution of the metal content of the ISM are roughly similar for the three SN II models (named A, B and C by Woosley & Weaver) but they are shifted to larger abundance in case of larger explosion energy. Curves for model B always lie between those of model A and C and will be omitted in the following studies.

2.3.4. Evolution of selected element abundances

As another improvement of our chemical evolution models we can calculate abundances [X/H] (cf. equation (5) in section 3) of a great variety of elements because appropriate input data are now available (as was pointed out in section 2.2). In **Fig. 5** we present abundances [X/H] for elements X = C, N, O, Mg, Al, Si, S, Ar, Ca, Cr, Mn, Fe, Ni and Zn in Sa galaxies.

At first sight all curves nearly have a similar shape (abundances increasing from high to low redshift) indicating that the enrichment history for all elements is roughly the same. But there are important differences in detail. Absolute element abundance values are very different (bearing in mind the log scale). Furthermore the gradients of the curves differ significantly in some parts reflecting the different production histories of various elements (i.e. SNII–, SNI– and intermediate mass star–products).

2.3.5. Influence of IMF and upper mass limit

Generally two different initial mass functions (IMF) are in use. Scalo's (1986) IMF is described in section 2.1 equation equation (1). Applying the same exponent $x = x_1 = x_2 = x_3 = 1.35$ for the whole mass range $m_{down} \leq m \leq m_{up}$ we recover Salpeter's (1955) IMF. Improved data, especially metallicity dependent yields by Woosley & Weaver (1995) are solely available up to $m_{up} = 40\,M_\odot$. To study the influence of the IMF and the upper mass limit $m_{up}$ on the results we therefore refer to earlier calculations (cf. Fritze 1995a and Fritze et al. 1995b) using only solar metallicity yields.

**Fig. 6** presents the time evolution of the global metallicity Z in Sa and Sd galaxies using Scalo and Salpeter IMFs and two different upper mass limits ($m_{up} = 40$ and $85\,M_\odot$). The range of metallicities observed in HII regions of nearby Sa to Sd galaxies (today = 15 Gyrs) by Oey & Kennicutt (1993) and Zaritsky et al.(1994) is indicated at the right edge of Fig. 6. We chose a Scalo IMF with $m_{up} = 40.0\,M_\odot$ (as described in section 2.1) throughout the paper.

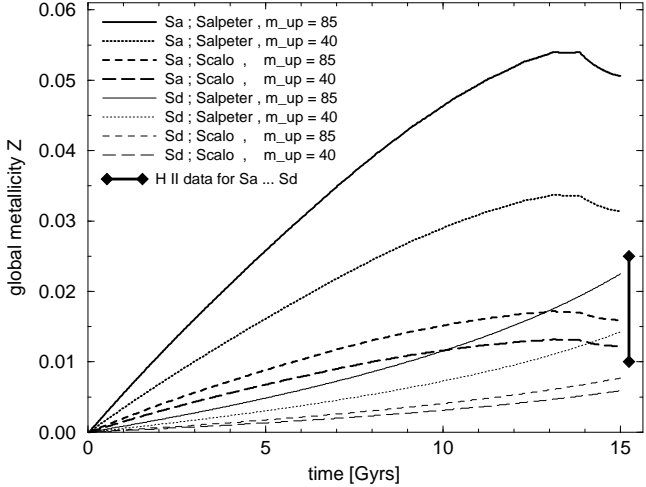

**Fig. 6.** Comparison of global metallicity Z for Sa and Sd galaxies using Salpeter and Scalo IMF with upper mass limit m_up = 85 and 40 solar masses, respectively. Metallicities of HII regions in nearby spiral galaxies are taken from Zaritsky *et al.* (1994) and Oey & Kennicutt (1993).

### 2.3.6. Comparison with H II region abundances of nearby spirals

Our simple one–zone models are assumed to give average ISM abundances. We therefore chose to compare them to observed nearby spiral H II region abundances as measured (or extrapolated from observed gradients) at $\sim R_e$. This is what Oey & Kennicutt (1993) call *characteristic abundances*. The range they give for Sa — Sb galaxies is $\sim 0.9 \, Z_\odot - 1.6 \, Z_\odot$ (cf. their Table 4). For Sbc through Sd spiral galaxies (Zaritsky *et al.* 1994) this characteristic abundance range extends downwards with average characteristic abundances (at 1 $R_e$) for Sd galaxies of about $0.5 \, Z_\odot$.

The compilation of Ferguson *et al.* (1998) confirms the radial gradients in galaxies out to large radii. In some cases they even find stronger gradients than those derived from the inner regions. Starting from observed spiral ISM abundances and abundance gradients and using a geometrical model Phillipps & Edmunds (1996) find that the average abundance encountered along an arbitrary line of sight through a present day spiral galaxy should be of the order $\sim \frac{1}{3} \, Z_\odot$.

It should be noted that oxygen abundances in H II regions may already be locally enhanced with respect to average ISM abundances as soon as the first supernovae explode among the stars that ionize the gas. We therefore decided to transform the observed H II region abundances [O/H] to a global metallicity Z for the comparison with our model results in Fig. 6.

### 2.3.7. Connection with DLA galaxies

It should be mentioned that for our 1-zone models it would not matter if at the highest redshifts the proto galaxies were not really assembled yet in one coherent structure but rather consisted of a set of subgalactic fragments that imprint their relative velocity differences on the structure of the DLA line profile. In this case, our model could be interpreted as describing the global SF and enrichment history of all the bits and pieces that are bound to later assemble into one present–day galaxy.

## 3. Observed abundances in DLA systems

We have compiled from the literature all available data on element abundances in DLA systems for comparison with results from our chemically consistent galaxy evolution models. Publications on abundance observations in DLA systems span more than one decade in time and the data experienced considerable improvement in quality and quantity in recent years. The total resulting compilation of element abundances is very inhomogeneous. We confine our comparison to abundances of Al, Ni, S, Fe, Si, Mn, Cr, and Zn, because for those eight elements enough observations are available. For all other elements included in our models the number of published abundance measurements in DLA systems is too small for a reasonable comparison. For our final compilation we take the following criteria into account.

- Many DLA systems (characterized by the redshift $z_{abs}$ of the absorption lines and the background QSO) are reported by more than one author. In these cases the most *reliable* data will be used.
- Different methods to determine column densities from observed absorption lines are in use: curve of growth analysis, profile fitting methods (e.g. VPFIT etc.), apparent optical depth method (Savage & Sembach 1991).
- The quality of data depends on instrumental capabilities. Resolution and signal to noise ratio of the spectrographs have been improved considerably in recent years using 4m class telescopes and the 10m Keck Telescope.
- All methods (mentioned above) used to derive column densities from observed absorption lines need oscillator strength f. Different f values are published in the literature and have been used by different observers. In column "ref$_f$" of Table 3 (cf. also footnote [b]) we list the references for oscillator strengths used for the DLA observations. In our final compilation we prepare a homogeneous database by referring all observed abundances to the oscillator strengths given by Morton (1991) since those are most widely used.
- Different solar element abundances are used by different authors to calculate abundances in the usual form

**Table 3.** Published element abundance measurements in DLA systems for Al, Ni, S, Fe, Si, Mn, Cr and Zn. Reference numbers for f-values (column $\mathrm{ref}_f$) and solar abundances (column $\mathrm{ref}_\odot$) are given in the footnotes.

| no. | authors | $\mathrm{ref}_\odot{}^a$ | $\mathrm{ref}_f{}^b$ | Al | Ni | S | Fe | Si | Mn | Cr | Zn |
|---|---|---|---|---|---|---|---|---|---|---|---|
| 1 | Blades, J.C., et al., 1985 | – | – | | | | 1 | 1 | | | |
| 2 | Boissé, P., et al., 1998 | 2 | – | | 4,2 | | 3,1 | 1 | 5,2 | 2,1 | 2,2 |
| 3 | Carswell, et al., 1991 | – | 5 | | | | | 1 | | | |
| 4 | Fan, Tytler 1994 | – | – | 2,2 | | | | 1 | | | |
| 5 | Hunstead, R.W., et al., 1987 | – | – | 1 | | | | 1 | | | |
| 6 | Hunstead, R.W., et al., 1986 | 7 | 4 | | | | | | | | |
| 7 | Lanzetta, K.M., Wolfe, A.M., Turnshek, D.A., 1989 | 1 | 5 | 1,1 | | | 1,1 | 1,1 | | | |
| 8 | Lu, L., et al., 1993 | – | 5 | 5 | | | 5 | 6 | | | |
| 9 | Lu, L., Savage, B.D., Tripp, T.M., Meyer, D.M., 1995 | 2 | 1/6/7 | 2,2 | 1 | | | 1 | | 1 | 1 |
| 10 | Lu, L., et al., 1996a | 2 | 1/6/8 | 8,2 15,10 | 3,1 | 16,14 | 12,10 | 7,6 | 15,11 | 11,4 | |
| 11 | Lu, L., et al., 1996b | 2 | 6/7 | 1 | 1 | | 1,1 | 1,1 | | | |
| 12 | Lu, L., Sargent, W.L.W., Barlow, T.A., 1997 | – | – | | | | 1,1 | | | | |
| 13 | Lu, L., Sargent, W.L.W., Barlow, T.A., 1998 | – | – | | | | 15,15 | | | | |
| 14 | Matteuchi, F., Molaro, P., Vladilo, G., 1997 | 2/6 | – | | | 2,1 | 2,2 | 1,1 | | | |
| 15 | Meyer, D.M., Roth, K.C., 1990 | 4 | 5 | | 7,5 | | | | | 4 | 3 |
| 16 | Meyer, D.M., Welty, D.E., York, D.G., 1989 | 4 | 5 | | 2,2 | | 1,1 | 1 | | | |
| 17 | Molaro, P., Centurión, M., Vladilo, G., 1997 | – | 6 | | | 1,1 | 1,1 | | 1 | | |
| 18 | Molaro, P., et al., 1996 | 2/6 | 6/8 | 1,1 | | | 1,1 | 1,1 | | | |
| 19 | Pettini, M., et al., 1994 | 1 | 5 | | | | | | | 9 | 5 |
| 20 | Pettini, M., Lipman, K., Hunstead, R.W., 1995 | 2 | – | 1,1 | 1 | 1,1 | 1,1 | 1,1 | | | |
| 21 | Pettini, M., et al., 1997 | 2 | 1 | | | | | | | | 15,14 |
| 22 | Prochaska, J.X., Wolfe, A.M., 1996 | 2 | 6 | | 1,1 | | 1,1 | 1,1 | | 1,1 | |
| 23 | Prochaska, J.X., Wolfe, A.M., 1997 | 2 | 1/6/8 | 1,1 | 2,1 | | 2,2 | 2,2 | | 1,1 | 2,1 |
| 24 | Rauch, M., et al., 1990 | 1 | – | | | | 1,1 | | | | |
| 25 | Turnshek, D.A., et al., 1989 | 2 | – | | | | 4 | 4 | | | |
| 26 | Vladilo, G., 1998 | 2/6 | – | | 7,7 | | 1,1 | | 1,1 | 7,7 | |
| 27 | Vladilo, et al., 1997 | – | – | | | | 1,1 | 1,1 | | | |
| 28 | Wolfe, A.M., et al., 1993 | – | – | | | | 4 | 3 | | | |
| 29 | Wolfe, A.M., et al., 1994 | 2 | – | | | 1,1 | | | | 1,1 | 1,1 |

Notes: [a] References on **solar element abundances**: 1: Aller, L.H., 1987; 2: Anders, E., Grevesse, N., 1989; 3: de Boer, K.S., Jura, M.A., Shull, J.M., 1987; 4: Grevesse, N., Anders, E., 1989; 5: Grevesse, N., Noels, A., 1993; 6: Hannaford, etal., 1992; 7: Withbroe, G.L., 1971;
[b] References on **oscillator strength**: 1: Bergeson, S.D., Lawler, J.E., 1993; 2: Kurucz, R.L., Peytremann, E., 1975; 3: Morton, D.C., Smith, W.H., 1973; 4: Morton, D.C., 1978; 5: Morton, D.C., York, D.G., Jenkins, E.B., 1988; 6: Morton, D.C., 1991; 7: Spitzer, L. Jr., Fitzpatrick, E., 1993; 8: Tripp, T.M., Lu, L., Savage, B.D., 1996

given in equation (5):

$$[X/H] := \log(X/H) - \log(X/H)|_\odot \quad (5)$$

where X denotes the number density $n(X)$ or column density $N(X)$ of element X measured from the absorption line and H denotes the hydrogen density, respectively. The symbol $|_\odot$ denotes solar values which are used as reference values (cf. Savage & Sembach, 1991). In column "$\mathrm{ref}_\odot$" of Table 3 (cf. also footnote [a]) we list the references to solar abundances used to calculate [X/H] by different authors. Our final compilation is homogeneously normalized to solar abundances published by Anders & Grevesse (1989) since they are most widely used.

Taking all these points into account we have compiled abundances in DLA systems for those eight elements which have the largest number of abundance determinations: Al, Ni, S, Fe, Si, Mn, Cr and Zn. According to the method used for abundance determination and the estimation of reliability of the data by the authors themselves we divide our sample into two classes of reliability: *reliable* and *less reliable*.

In **Table 3** we list 29 papers, each reports at least one abundance for any of our eight elements. The quan-tity of data in each paper is given in the columns labeled with respective elements. The number of *reliable* data for any element is given by the second number separated by a comma. For instance, Boissé *et al.* (1998) report Fe abundances for three DLA systems, one of these is *reliable* and two are *less reliable* measurements. Obviously the main share of data comes from about five papers published during the past five years (their running number is printed heavy).

## 4. Results and discussion

### 4.1. Comparison of model calculations with DLA observations

In **Fig. 7. a − h** we present abundances in DLA systems for eight elements Fe, Si, Zn, Cr, Ni, S, Al, and Mn. The data are compiled from the 29 papers listed in Table 3. *Reliable* and *less reliable* data are marked with filled and open circles, respectively. For several abundances the authors report only upper or lower limits which are indicated by open triangles pointing downwards or upwards, respectively. Those data, of course, are classified as *less reliable*.

In Fig. 7.a) to 7.d) some pairs of data points representing abundances measured for the same DLA system by two different authors are connected by heavy vertical lines. They show how abundance measurements by different authors still can differ and give us a means to estimate the observational errors. In Fig. 7.a) the DLA system at $z \sim 0.7$ has two *reliable* values differing by about 0.2 dex in [Fe/H] and it lies just beneath the Sd curve. The second set of independent measurements in Fig. 7.a) belongs to the DLA system with $z_{abs} \sim 2.15$. The lower value is indicated as a *lower limit* which is in agreement with the larger *reliable* value. Generally *reliable* measurements of different authors agree to within $\sim 0.3$ dex. Errors reported by a few authors for some (mostly *reliable*) data are indicated as weak vertical lines in Fig. 7 and likewise are about $\pm 0.3$ dex.

The model curves in Fig. 7 represent the redshift evolution of Sa and Sd type galaxies as calculated from our chemically consistent chemical evolution models (heavy lines) and models using exclusively input physics of solar abundance (weak lines). In the legend they are indicated as "(chem.cons.)" and "(solar)", respectively. Sb and Sc galaxies are omitted to avoid overcrowding of the figures. Their curves always lie between those of Sa and Sd as demonstrated in Fig. 3. Furthermore, for the chemically consistent calculations, we also present the redshift evolution of element abundances using Woosley & Weaver's SN II yields from their model C, which has larger explosion energies (cf. section 2.2), indicated as "model C" in Fig. 7. Model B curves are omitted because they always fall between our heavy lines (which use yields from model A) and the curves for model C (cf. Fig. 4, section 2.3).

### 4.2. General implications for the models

For all eight elements under consideration almost all data points lie between our chemically consistent model curves for Sa and Sd galaxies. We find particularly good agreement between models and observations for the elements Zn and Ni where many observations are available and for Al, Mn and S with a smaller number of (*reliably*) observations. Having in mind the fact that Sb and Sc models lie within the region outlined by the Sa and Sd curves (cf. Fig. 3 in section 2.3) we can establish nearly perfect conformity between element abundances observed in DLA systems and our model calculations for spiral galaxies spanning the whole redshift range from 0 to 4.5. And since our models for $z = 0$ agree well with observed average ISM abundances of nearby spiral galaxies it is clear that **DLA galaxies may well evolve into the full range of present–day spiral galaxies**, although we cannot exclude the possibility that a few DLA systems might be LSB galaxies or (starbursting) dwarfs.

#### 4.2.1. Implications for chemically consistent models

Differences for chemical consistent vs. pure solar metallicity models on the one hand and different SN II explosion energies (model A vs. model C by Woosley & Weaver 1995) on the other hand are small compared to differences due to the variation of the star formation rate characterizing our spectral galaxy types Sa ... Sd. Consequently none of the models can be excluded or is clearly favored by comparison with the DLA data. They all are in conformity with the observations. **The range of $\sim 1.5 \cdots 2$ dex of element abundance in DLA systems at any given redshift is naturally explained by the range of star formation rates among early to late type spirals** as outlined in section 2.1.

#### 4.2.2. Scatter in observational data

Additionally, some observational scatter is expected in DLA element abundances: The column densities observed in DLAs could depend upon the (unknown) impact parameter (if abundance gradients already exist in (proto-) galactic disks at high redshift), on inclination effects, and on local inhomogeneities along the line of sight (cf. the different abundances determined for the cold and warm disk component of our Galaxy (Savage & Sembach 1996)).

#### 4.2.3. Observed abundances exceeding our Sa model

In case of Fe, Si, Cr, Zn and S four *reliable* abundance measurements clearly exceed the values calculated for our Sa model. The corresponding DLAs are listed in **Table 4** and the elements with abundance in excess of our Sa model prediction are marked with "X" ("O" indicates conformity of observations with models and for "–" no observations are available). The small error bars attached to the data points indicate that these deviations cannot be due to observational errors.

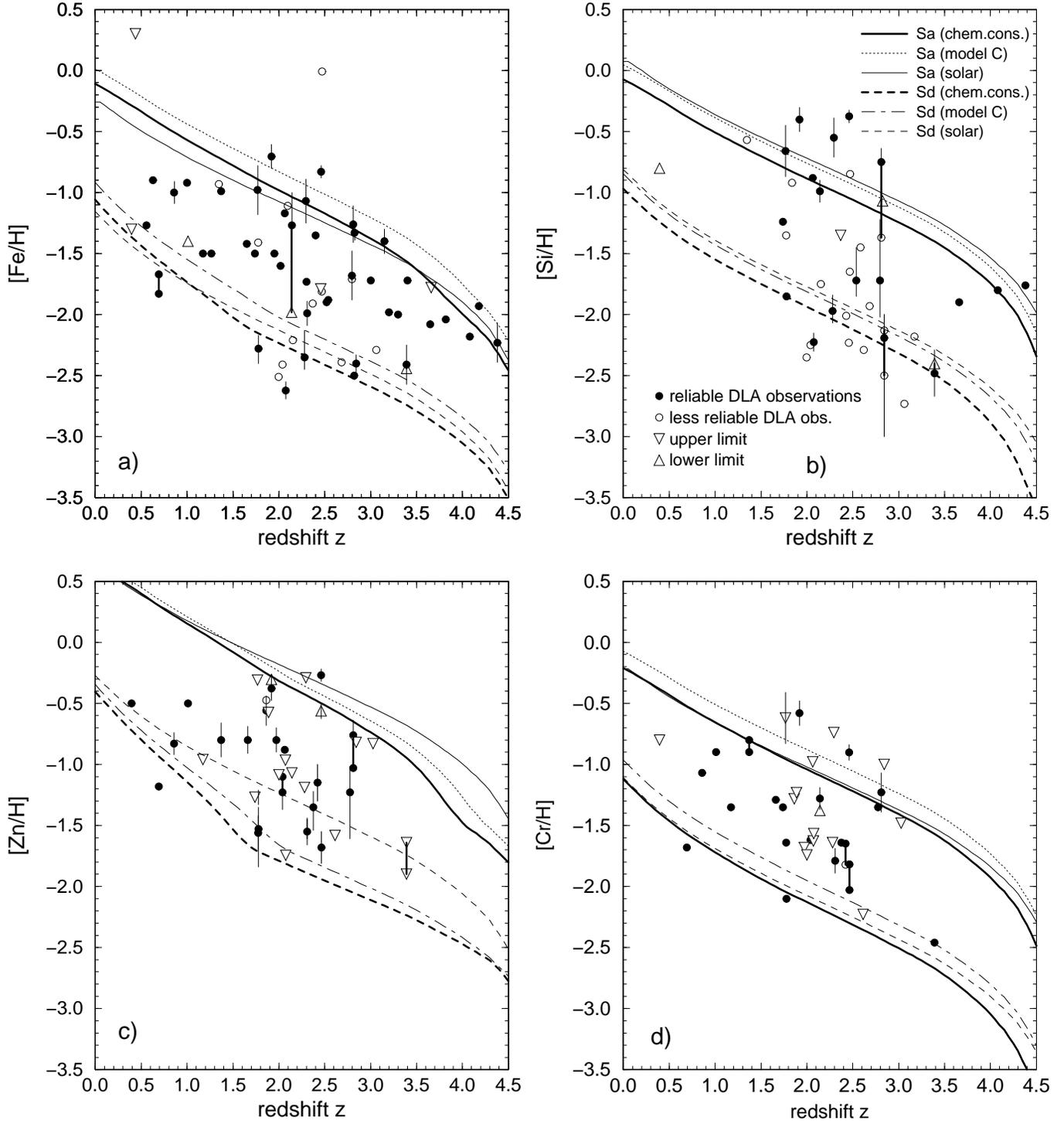

**Fig. 7.** redshift dependence of element abundances a) [Fe/H], b) [Si/H], c) [Zn/H] and d) [Cr/H] for observed DLA systems and various Sa and Sd model galaxies.

It is seen that in two of the four cases (QSO 0216+0803 and 0528-2505) only the typical SNII elements Si and S show abundances higher than those of our Sa model whereas the iron group elements Fe, Zn and Cr which have important SNI contributions are **not** enhanced. For these two DLAs a temporarily enhanced SFR or a small star

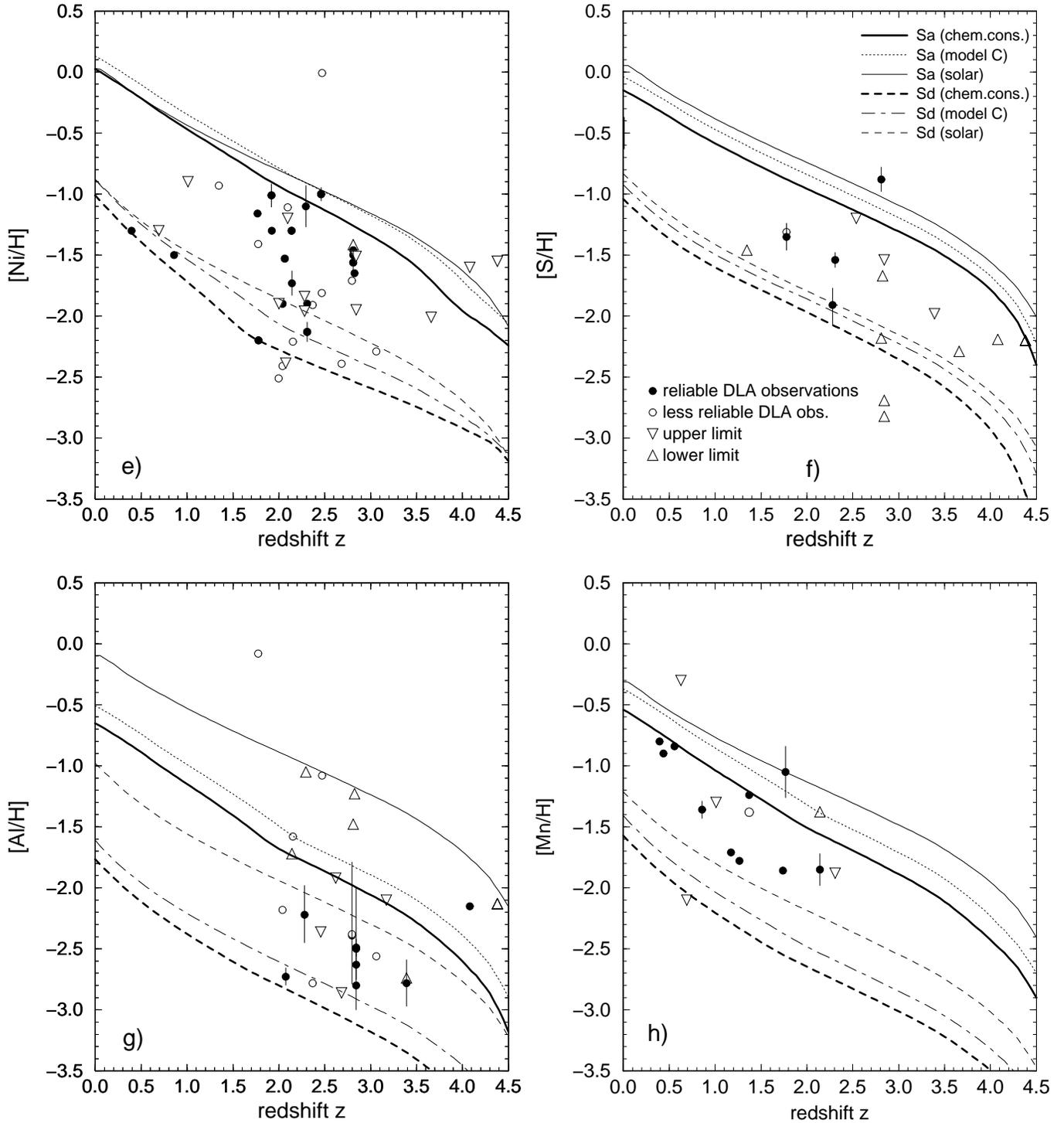

**Fig. 7. continued:** redshift dependence of element abundances e) [Ni/H], f) [S/H], g) [Al/H] and h) [Mn/H] for observed DLA systems and various Sa and Sd model galaxies.

burst in an early type spiral galaxy could easily explain their abundance pattern.

While, of course, we cannot exclude that some dwarf or LSB galaxies may also be present among the DLA absorber sample, our models indicate that the bulk of DLA abundances and, in particular, their redshift evolution, are consistent with them being normal spiral galaxy progenitors.

For the other two systems in QSO 2206-199 and 0201+365 both the SNII product Si as well as elements with important SNI contributions like Fe have abundances higher than those of our Sa model. For these we conclude that the enhancement probably is not due to a temporarily enhanced SFR or star burst but rather to a characteristic timescale of star formation shorter than the $t_* = 2$ Gyr adopted for our Sa model. Our models cannot tell if these two DLA systems are the progenitors of S0 galaxies or of a bulge component, which both are believed to form the bulk of their stars on a short timescale. Alternatively the high abundances might result from a very early star formation enhancement, a formation of these systems at z > 5 or a local overabundance where the line of sight is passing through the disk.

**Table 4.** DLA systems with element abundances in excess of our Sa model are indicated with "X", and "x" for *less reliable* data. "O" indicates conformity of observations with models and for "–" no observations are available.)

| QSO name | $z_{abs}$ | Fe | Si | Cr | Zn | S |
|---|---|---|---|---|---|---|
| 2206–199 | 1.92 | X | X | X | O | – |
| 0216+0803 | 2.29 | O | X | x | x | – |
| 0201+365 | 2.46 | X | X | X | X | – |
| 0528-2505 | 2.81 | O | X | O | O | X |

### 4.2.4. Chemically consistent versus purely solar models

In the case of Zn, Ni and Al we find a significant difference between chemically consistent models and the comparison models using solar metallicity stellar yields only. In Fig. 7.c) and 7.e), i.e. for Zn and Ni, a considerable number of observations lie below the Sd curve (dashed line) of the solar metallicity model (weak line) but within the region outlined by our chemically consistent models (heavy line). In the case of Al ( Fig. 7.g), almost all data points lie between the chemically consistent evolution models (heavy line), but most of the observed abundances drop below the Sd curve of the solar metallicity model.

To conclude: **For some elements the chemically consistent models do not differ very much from the solar metallicity ones, for other elements they do yield significantly lower abundances, and in these cases the chemically consistent calculations do fit the observations much better than the solar metallicity models.**

### 4.3. Discussion of model parameters IMF and SFR

In **Fig. 8** we present observed Fe abundances [Fe/H] in DLA galaxies together with model calculations using different IMFs (Scalo vs. Salpeter) and for Sd galaxies different star formation rates (SFR). Resulting curves using a Salpeter IMF are plotted with weak lines whereas those for the Scalo IMF are displayed by heavy lines as before. We find that abundances calculated with a Salpeter IMF are larger than those resulting from models using a Scalo IMF (cf. Fig. 6 in section 2.3.5). Some large [Fe/H] values observed in DLA systems can now be reached with the Salpeter IMF but then, in turn, some low abundance data fall outside the region outlined by our models. This drawback can be compensated by lowering the star formation rate of Sd galaxies. Because the SFR of Sd galaxies is constant this can be done without changing the photometric properties of the galaxies.

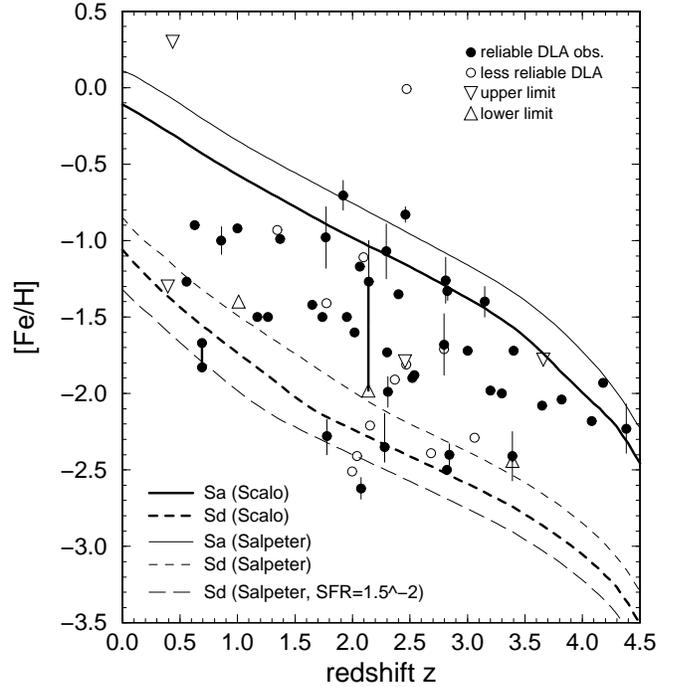

**Fig. 8.** redshift dependence of Fe abundance in observed DLA galaxies along with model calculations with different model parameters: Scalo vs. Salpeter IMF for Sa and Sd galaxies and two different constant star formation rates SFR_1= $3.5 \cdot 10^{-2}$ and SFR_2= $1.5 \cdot 10^{-2}$ for Sd in case of Salpeter IMF.

It should be mentioned that while the SFHs of nearby galaxies seem to be very homogeneous for early type spirals the scatter around the average SFH used in our models significantly increases towards later types. This can be seen e.g. in the very small range of $H_\alpha$ equivalent widths or colors among Sa galaxies as opposed to the much larger scatter both in $H_\alpha$ equivalent widths and colors seen among local Sd galaxies (cf. Kennicutt & Kent 1983, Buta et al. 1994).

Very small SFRs are characteristic for *low surface brightness galaxies* (LSBs) and hence we conclude that a few observed DLA systems with particularly low abundances could be either late type spirals with a lower than average SFR or else LSB galaxies. As discussed in the previous subsection an extremely low element abundance

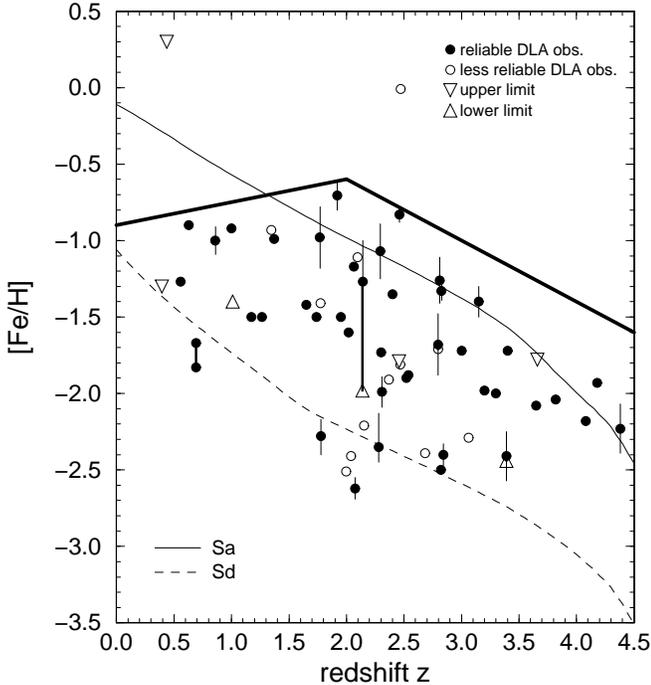

**Fig. 9.** This reduced version of Fig. 7.a) shows the upper envelope (heavy line) to the observed Fe abundances in DLAs and our chemically consistent evolution models for Sa and Sd galaxies.

measured in DLA systems could also be due to a large (unknown) impact parameter.

## 5. Implications for DLA systems

### 5.1. DLAs at low redshift: abundances versus gas content

In **Fig. 9** we present a reduced version of Fig. 7.a) to show that the upper envelope to the observed DLA abundances (heavy line) clearly increases from $z \sim 4.5$ towards $z \sim 2$ but declines or remains constant from $z \sim 2$ to $z \sim 0$. This behavior is seen for all eight elements investigated in Fig. 7. For $z \sim 4.5$ through $z \sim 2$ abundance data for all these elements completely cover the range between our models for Sa through Sd galaxies. For $z < 2$ the situation changes. The lower the redshift of DLA absorbers the closer do their observed abundances fall to our early type spiral models. It looks as if at $z < 2$ Sa disks would no longer have large enough cross sections at the high HI column densities required for damped Ly$_\alpha$ absorption to appear in DLA samples. Virtually all of the $z < 1$ abundance data fall close to our models for Sc or Sd galaxies. In our simple 1–zone galaxy models we do not have any information about gas densities but only about the total gas content.

Interestingly, this global gas content for our Sa model drops from more than 50 % (of the total mass) at $z \sim 2$ to $\sim 40$ % at $z \sim 1.5$ while that of our Sd model still is more than 60 % at $z \sim 0$. So, neglecting any density structure of the HI disk which is not included in our modelling the comparison between observed abundance data and the enrichment evolution of our models suggests that as the global gas content falls below 50 % galaxies drop out of the DLA absorber sample. From their efforts to optically identify DLA systems Steidel et al. (1995) suspect that there is "a selection effect against luminous spiral galaxies (like our Galaxy) for moderate redshift DLA systems" (see also Steidel et al.1997). It would be interesting to check our prediction for low-z DLA galaxy types with HST.

### 5.2. Implications for optical identifications

The comparison of our spiral models with observed DLA abundances suggests that the bulk of the DLAs are the high redshift progenitors of present–day spiral galaxies from Sa through Sd. This has implications for the possibilities of optical detections.

Spiral galaxies at $z = 0$ have luminosities in the range from $M_B = -17.7$ for Sd galaxies to $M_B = -19.7$ for Sa galaxies, when using the mean values for these types in the Virgo cluster as given by Sandage et al. (1985). In any case, typical spirals are significantly fainter than $L^*$. Moreover we argued in section 5.1 that the brighter early type spirals seem to drop out of the DLA sample towards low redshift due to their gas content becoming too scarce. This leads us to expect that **low redshift DLA samples should be dominated by the particularly faint gas–rich late–type spiral (or even irregular or LSB) galaxies**. Hence we do expect early type spirals to be among the DLA absorbers only for redshift larger than $z \sim 1.5$.

With evolutionary and cosmological corrections calculated from our chemically consistent spectrophotometric evolutionary synthesis code (Möller et al. 1998) we find that the typical luminosity of Sa galaxies increases up to $M_B \sim -21 \cdots -21.7$ in the range $z = 2 \cdots 3$. Taking into account bolometric distance moduli for $z = 2$ and 3 and using cosmological parameters $H_0 = 50$ km s$^{-1}$ Mpc$^{-1}$ and $\Omega_0 = 1$ this yields apparent magnitudes of about $m_B \sim 24.5 \cdots 25$ which are close to the detection limit. Later galaxy types are fainter, e.g. an average Sd galaxy at $z \sim 2.5$ has $M_B \sim -18$ and $m_B \sim 28.5$, at $z \sim 1.5$ it has $M_B \sim -17.6$ and $m_B \sim 27.5$ and at $z \sim 0.5$ an average Sd galaxy has $M_B \sim -17.1$ and $m_B \sim 25.5$. In terms of $\mathcal{R}$-magnitudes we expect the average early type spiral Sa to have $\mathcal{R} \sim 25.8$ at $z \sim 2$ and $\mathcal{R} \geq 26$ at $z \geq 2.5$. This explains why deep surveys did **not** detect DLAs at $z \sim 3$ down to $\mathcal{R} \sim 25.5$ (cf. Steidel et al. 1998). In the K–band our models predict $K \sim 21.4$ for Sa and $K \sim 26$ for Sd galaxies at $z \sim 2$. This also makes us understand that on deep K images of ten QSOs with DLA systems in the redshift range $1.5 \leq z \leq 2.5$ investigated by Aragon–Salamanca et al. (1996) only two candidates with $K \sim 20$ (i.e. from the bright end of the spiral galaxy luminosity function) have been detected. Since our chemical evolution models suggest that the brighter early type galaxies should drop out of the DLA sample due to too scarce gas content towards low redshift we expect optical identifications of low redshift DLA systems not to be easier: an average late-type spiral at z $\sim$ 0.5 is expected to have B $\sim$ 25.5, $\mathcal{R}$ $\sim$ 25 and K $\geq$ 21.5. The galaxies identified by Steidel *et al.*(1994 and 1995) in the fields of 3C 286 (Q1328+307), PKS 1229-021 and PKS 0454+039 as candidates for DLA absorbers at $z_{abs}$ = 0.6922, $z_{abs}$ = 0.3950 and 0.7568 and $z_{abs}$ = 0.8596, respectively, indeed have typical luminosities between $M_B$ $\sim$ $-19$ and $-19.5$ of late type spiral galaxies. We conclude that many DLA candidates are out of reach for present day imaging capabilities and hence for optical identification.

*5.3. Possible implications for abundance ratios*

If our result that DLA absorbers might well be associated with the progenitors of normal present–day spiral galaxies of all types from Sa through Sd were confirmed by high resolution imaging observations, a direct by–product of our models could be the ISM abundance ratios of various elements and their evolution with redshift. To the extent that observed Milky Way disk star abundance ratios directly reflect the ISM abundance ratios at birth of those stars we would then expect a disk–like rather than a halo–like abundance pattern for DLA systems (cf. the controversy on halo– vs. disk–like DLA abundance patterns between Lu *et al.* (1996) and Kulkarni *et al.* (1997)). Due to uncertainties in the stellar input yields, a rather small number of observed DLA abundance ratios available at present, and a deficient knowledge of how to correct for the effect of selective dust condensation, however, we estimate it premature to base any conclusion on such a comparison at present.

## 6. Conclusions

We have compiled from the literature all available data on element abundances in DLA systems measured over the last decade, referred them all to one set of oscillator strength and solar reference values and subdivided them into *reliable* and *less reliable* data.

We extended our chemical galaxy evolution model to calculate abundances for altogether 16 different elements in spiral galaxies in a chemically consistent way, i.e. accounting for the steadily increasing initial metallicities of successive generations of stars by using input data bases (stellar yields, evolutionary tracks, lifetimes and remnant masses) for 5 different metallicities from Z = 0 $\cdots$ Z = $Z_\odot$. Dynamical effects, however, are not included in our simple 1–zone models.

A detailed comparison of model results with observations for eight different element abundances yields the following main results:

- The conformity between observed element abundances in DLA systems and those calculated from our models for spiral galaxies spanning the whole redshift range $0 \leq z \leq 4.5$ indicates that DLA galaxies may well evolve into the full range of present–day spiral galaxies.
- Without any adjustments and only using SFHs that proved successful for the spectrophotometric description of spiral galaxies from zero to high redshift our models successfully bridge the gap between abundances observed in high redshift DLA absorbers and the H II regions of present–day normal spiral galaxies Sa through Sd.
- The slow redshift evolution of DLA abundances (compared to halo CIV systems) is a consequence of the relatively long timescales for star formation in disk galaxies.
- The large scatter observed in element abundances in DLAs of similar redshift is naturally explained by the range of star formation rates at any redshift between early and late type spiral galaxies.
- The few observations exceeding the abundances calculated for our Sa model can be explained either by a temporarily enhanced SFR or a small starburst in early type spirals if only SNII products are enhanced or by a shorter characteristic timescale for star formation, very early star formation enhancement or a local overabundance along the line of sight if both SNI and SNII products are enhanced.
- Comparison of our chemically consistent models with models using only solar metallicity input physics shows that differences in the redshift evolution are small for some elements but large for others. For the elements Zn, Ni and Al with large differences the chemically consistent models provide a significantly better agreement with observed DLA abundances.
- Using a Salpeter IMF instead of Scalo yields larger element abundances throughout and many data points fall below the curve of the Sd model. This could be compensated by using a lower SFR and means that more DLAs could be LSB galaxies.

Our comparison of element abundances observed in DLA systems with those resulting from our chemically consistent galaxy evolution models has important implications for the nature of low redshift DLAs and the possibility of optical identification of DLAs over the whole redshift range.

- The upper envelope to observed DLA abundances increases from z $\sim$ 4.5 to z $\sim$ 2 and decreases to smaller redshifts whereas our Sa model abundances increase steadily. This leads to the conclusion that Sa galaxies at low redshift may not have gas at sufficient HI column densities over large enough cross sections to cause damped Ly$_\alpha$ absorption, whereas Sd galaxies appear as DLA systems down to z $\sim$ 0. This can explain why much less DLA systems have been found at low redshift in the HST key project QSO Absorption Lines (Bahcall *et al.* 1993) than expected from their high redshift frequency.

– The comparison between our models and observations suggests that DLA systems could be the progenitors of Sa to Sd type galaxies with intrinsically faint late type spirals dominating at low redshift. Estimates from our chemically consistent spectrophotometric evolution models – including evolutionary and cosmological corrections – predict comparable luminosities for the brighter early type spiral DLA galaxies at $z \sim 2 \cdots 3$ and for the intrinsically fainter late type spirals expected to dominate DLA samples at $z \sim 0.5$: $B \sim 25.5$, $\mathcal{R} \sim 25$ and $K \sim 21.5$.

*Acknowledgements.* This work was supported by Deutsche Forschungsgemeinschaft grant Fr 325/40-1. We thank Stan E. Woosley for providing us with the data files of all tables published in ApJS 101, 181 (1995) in machine readable form. Detailed and helpful comments from an anonymous referee are gratefully acknowledged.

**Table 7.** Stellar lifetimes $\tau(m_*)$ given in Gyrs = $10^9$ yrs for two different initial metallicities. $\tau_1$: Z = $Z_\odot$ and $\tau_2$: Z = 0.05 $Z_\odot$

| $m_*$ | $\tau_1$ | $\tau_2$ | $m_*$ | $\tau_1$ | $\tau_2$ | $m_*$ | $\tau_1$ | $\tau_2$ |
|---|---|---|---|---|---|---|---|---|
| .15 | 20.0 | 20.0 | 1.7 | 1.856 | 1.305 | 15 | $1.27^{-2}$ | $1.45^{-2}$ |
| .30 | 20.0 | 20.0 | 2.0 | 1.412 | 1.029 | 18 | $1.02^{-2}$ | $1.18^{-2}$ |
| .45 | 20.0 | 20.0 | 2.5 | 0.757 | 0.571 | 20 | $8.96^{-3}$ | $1.03^{-2}$ |
| .60 | 20.0 | 20.0 | 3.0 | 0.441 | 0.341 | 22 | $8.20^{-3}$ | $9.20^{-3}$ |
| .70 | 20.0 | 20.0 | 4.0 | 0.194 | 0.166 | 25 | $7.05^{-3}$ | $7.85^{-3}$ |
| .80 | 15.2 | 15.2 | 5.0 | 0.108 | 0.100 | 30 | $6.20^{-3}$ | $7.00^{-3}$ |
| .90 | 13.0 | 10.0 | 7.0 | 0.048 | 0.050 | 35 | $5.55^{-3}$ | $6.20^{-3}$ |
| 1.00 | 11.4 | 6.9 | 8.0 | 0.029 | 0.032 | 40 | $4.79^{-3}$ | $5.34^{-3}$ |
| 1.30 | 5.0 | 3.5 | 12.0 | 0.018 | 0.020 | | | |

**Table 8.** SN II remnants of various progenitor masses $m_*$, five different initial metallicities $Z$ and three explosion models (A, B and C) from TABLE 3 in Woosley & Weaver (1995)

| $m_*$ | 1.0 $Z_\odot$ | 0.1 $Z_\odot$ | 0.01 $Z_\odot$ | $10^{-4}$ $Z_\odot$ | 0.0 $Z_\odot$ |
|---|---|---|---|---|---|
| 12 | 1.32 | 1.38 | 1.40 | 1.28 | 1.35 |
| 13 | 1.46 | 1.31 | 1.44 | 1.44 | 1.28 |
| 15 | 1.43 | 1.49 | 1.56 | 1.63 | 1.53 |
| 18 | 1.76 | 1.69 | 1.58 | 1.61 | 3.40 |
| 20 | 2.06 | 1.97 | 1.98 | 1.97 | 4.12 |
| 22 | 2.02 | 2.12 | 2.04 | 2.01 | 1.49 |
| 25 | 2.07 | 1.99 | 1.87 | 1.87 | 6.36 |
| 30 | 4.24 | 2.76 | 3.22 | 2.89 | 8.17 |
| $30^*$ | 1.94 | 2.01 | 2.21 | 2.08 | 1.54 |
| 35 | 7.38 | 6.69 | 5.41 | 10.10 | 12.80 |
| $35^*$ | 3.86 | 3.39 | 2.42 | 3.03 | 7.62 |
| $35^{**}$ | 2.03 | 2.02 | 1.96 | 1.97 | 1.85 |
| 40 | 10.34 | 9.13 | 9.08 | 13.70 | 16.60 |
| $40^*$ | 5.45 | 4.45 | 4.42 | 4.09 | 12.20 |
| $40^{**}$ | 1.98 | 2.01 | 2.02 | 2.03 | 1.99 |

Notes: $^*$model B, $^{**}$model C

## Appendix : Tables for electronic publishing

**Table 5.** SN II yields calculated by Woosley & Weaver (1995) for models A, B and C (see text) and two different metallicities $Z = Z_\odot$ and $Z = 0.01\,Z_\odot$. In the case of $m_* = 25\,m_\odot$ yields calculated from model A for all five different metallicities $Z = Z_\odot$, $Z = 0.1\,Z_\odot$, $Z = 0.01\,Z_\odot$, $Z = 10^{-4}\,Z_\odot$ and $Z = 0\,Z_\odot$ are given. $E_{tot}$ is the total ejected mass as given by Woosley & Weaver (1995). $\Delta$ indicates the difference (in %) between $E_{tot}$ and the sum of ejecta given in this table for elements H, He, ..., Mn.

| $m_*$ | H | He | C | N | O | Mg | Al | Si | S | Ar | Ca | Fe | Ni | Zn | Cr | Mn | $E_{tot}$ | $\Delta$ |
|---|---|---|---|---|---|---|---|---|---|---|---|---|---|---|---|---|---|---|
| $Z = Z_\odot$ | | | | | | | | | | | | | | | | | | |
| 12 | 5.96 | 4.11 | .082 | .036 | .219 | .011 | .001 | .093 | .079 | .027 | .015 | .059 | .010 | .0012 | .0010 | .0005 | 10.74 | .33 |
| 13 | 6.32 | 4.51 | .115 | .047 | .275 | .023 | .002 | .062 | .028 | .005 | .004 | .151 | .018 | .0023 | .0014 | .0007 | 11.62 | .49 |
| 15 | 6.98 | 5.24 | .162 | .054 | .685 | .040 | .005 | .116 | .066 | .015 | .011 | .138 | .012 | .0014 | .0021 | .0011 | 13.65 | .90 |
| 18 | 7.89 | 6.28 | .249 | .057 | 1.146 | .077 | .010 | .146 | .059 | .010 | .007 | .093 | .006 | .0005 | .0023 | .0012 | 16.34 | 1.87 |
| 20 | 8.24 | 6.72 | .214 | .060 | 1.953 | .049 | .004 | .300 | .176 | .037 | .015 | .120 | .008 | .0008 | .0036 | .0017 | 18.04 | .76 |
| 22 | 8.79 | 7.51 | .242 | .067 | 2.385 | .062 | .008 | .382 | .190 | .036 | .018 | .241 | .019 | .0019 | .0044 | .0021 | 20.09 | .64 |
| 25 | 9.40 | 8.64 | .324 | .080 | 3.254 | .162 | .025 | .339 | .149 | .027 | .017 | .169 | .013 | .0015 | .0031 | .0031 | 23.07 | 2.01 |
| 30 | 10.50 | 10.40 | .288 | .104 | 3.652 | .270 | .040 | .141 | .016 | .002 | .001 | .029 | .004 | .0015 | .0004 | .0003 | 25.90 | 1.74 |
| 30* | 10.50 | 10.40 | .292 | .104 | 4.882 | .346 | .051 | .384 | .108 | .019 | .013 | .481 | .053 | .0042 | .0047 | .0026 | 28.16 | 1.83 |
| 35 | 11.50 | 11.90 | .303 | .125 | 3.072 | .188 | .030 | .059 | .012 | .002 | .002 | .032 | .004 | .0013 | .0005 | .0004 | 27.82 | 2.12 |
| 35* | 11.50 | 11.90 | .322 | .125 | 5.822 | .392 | .064 | .166 | .019 | .003 | .002 | .032 | .006 | .0026 | .0005 | .0004 | 31.33 | 3.11 |
| 35** | 11.50 | 12.00 | .322 | .125 | 6.362 | .407 | .066 | .494 | .179 | .032 | .024 | .625 | .076 | .0059 | .0062 | .0060 | 33.23 | 3.01 |
| 40 | 11.10 | 13.00 | .315 | .141 | 2.361 | .124 | .021 | .031 | .011 | .002 | .002 | .032 | .003 | .0010 | .0005 | .0004 | 27.67 | 1.90 |
| 40* | 11.10 | 13.00 | .365 | .141 | 6.031 | .363 | .065 | .080 | .013 | .003 | .002 | .033 | .006 | .0027 | .0005 | .0004 | 32.55 | 4.13 |
| 40** | 11.10 | 13.00 | .370 | .141 | 7.631 | .449 | .077 | .621 | .248 | .043 | .032 | .760 | .077 | .0076 | .0100 | .0051 | 35.98 | 3.91 |
| $Z = 0.1\,Z_\odot$ | | | | | | | | | | | | | | | | | | |
| 12 | 6.26 | 3.91 | .089 | .003 | .145 | .006 | .000 | .030 | .014 | .003 | .003 | .178 | .016 | .0022 | .0012 | .0004 | 10.68 | .17 |
| 13 | 6.67 | 4.30 | .109 | .004 | .290 | .018 | .001 | .066 | .032 | .006 | .005 | .186 | .021 | .0026 | .0018 | .0009 | 11.76 | .39 |
| 15 | 7.42 | 5.04 | .142 | .005 | .555 | .023 | .001 | .072 | .034 | .007 | .006 | .197 | .010 | .0018 | .0018 | .0006 | 13.60 | .60 |
| 18 | 8.36 | 6.18 | .225 | .006 | .995 | .080 | .004 | .130 | .049 | .009 | .008 | .140 | .005 | .0009 | .0023 | .0007 | 16.41 | 1.31 |
| 20 | 8.95 | 6.70 | .246 | .006 | 1.520 | .054 | .003 | .234 | .131 | .026 | .019 | .126 | .003 | .0002 | .0038 | .0008 | 18.13 | .59 |
| 22 | 9.44 | 7.42 | .232 | .007 | 2.120 | .019 | .002 | .329 | .189 | .042 | .023 | .128 | .003 | .0001 | .0045 | .0009 | 20.00 | .20 |
| 25 | 10.30 | 8.60 | .267 | .009 | 2.900 | .032 | .003 | .399 | .223 | .054 | .022 | .207 | .011 | .0005 | .0036 | .0010 | 23.09 | .24 |
| 30 | 11.40 | 10.50 | .304 | .011 | 4.110 | .328 | .017 | .079 | .003 | .000 | .000 | .002 | .000 | .0001 | .0000 | .0000 | 27.40 | 2.36 |
| 30* | 11.40 | 10.50 | .304 | .011 | 4.420 | .344 | .018 | .214 | .065 | .013 | .011 | .185 | .009 | .0003 | .0028 | .0010 | 28.14 | 2.28 |
| 35 | 12.40 | 12.00 | .319 | .013 | 3.100 | .152 | .008 | .027 | .002 | .000 | .000 | .002 | .000 | .0001 | .0000 | .0000 | 28.53 | 1.77 |
| 35* | 12.40 | 12.00 | .334 | .013 | 5.780 | .355 | .019 | .115 | .006 | .001 | .000 | .002 | .000 | .0001 | .0000 | .0000 | 31.84 | 2.56 |
| 35** | 12.40 | 12.10 | .333 | .013 | 5.980 | .358 | .020 | .366 | .149 | .029 | .027 | .616 | .041 | .0024 | .0087 | .0022 | 33.26 | 2.45 |
| 40 | 13.30 | 14.00 | .337 | .016 | 2.720 | .084 | .005 | .009 | .002 | .000 | .000 | .002 | .000 | .0001 | .0000 | .0000 | 31.09 | 1.98 |
| 40* | 13.30 | 14.00 | .399 | .016 | 6.250 | .282 | .016 | .030 | .002 | .000 | .000 | .002 | .000 | .0002 | .0000 | .0000 | 35.78 | 4.14 |
| 40** | 13.30 | 14.00 | .400 | .016 | 7.150 | .342 | .019 | .426 | .197 | .039 | .035 | .680 | .049 | .0019 | .0068 | .0055 | 38.16 | 3.92 |
| $Z = 0.01\,Z_\odot$ | | | | | | | | | | | | | | | | | | |
| 12 | 6.30 | 3.90 | .090 | .000 | .142 | .006 | .000 | .031 | .014 | .003 | .003 | .152 | .012 | .0016 | .0009 | .0003 | 10.67 | .13 |
| 13 | 6.71 | 4.28 | .109 | .000 | .210 | .011 | .000 | .042 | .022 | .005 | .005 | .191 | .012 | .0020 | .0019 | .0004 | 11.63 | .25 |
| 15 | 7.49 | 5.09 | .149 | .001 | .423 | .022 | .001 | .074 | .035 | .007 | .006 | .172 | .011 | .0013 | .0020 | .0005 | 13.53 | .34 |
| 18 | 8.57 | 6.25 | .194 | .001 | .952 | .037 | .001 | .101 | .056 | .012 | .010 | .136 | .004 | .0005 | .0025 | .0006 | 16.52 | 1.16 |
| 20 | 9.07 | 6.62 | .208 | .001 | 1.620 | .017 | .001 | .245 | .150 | .038 | .019 | .075 | .001 | .0000 | .0035 | .0006 | 18.13 | .33 |
| 22 | 9.67 | 7.50 | .248 | .001 | 1.940 | .058 | .003 | .314 | .150 | .029 | .019 | .100 | .002 | .0000 | .0040 | .0006 | 20.08 | .20 |
| 25 | 10.40 | 8.76 | .279 | .001 | 2.830 | .040 | .003 | .388 | .208 | .051 | .020 | .188 | .011 | .0004 | .0028 | .0008 | 23.24 | .24 |
| 30 | 11.60 | 10.30 | .315 | .001 | 3.980 | .257 | .009 | .119 | .003 | .000 | .000 | .000 | .000 | .0000 | .0000 | .0000 | 26.95 | 1.36 |
| 30* | 11.60 | 10.30 | .315 | .001 | 4.450 | .277 | .010 | .312 | .099 | .019 | .018 | .177 | .006 | .0002 | .0034 | .0013 | 27.95 | 1.29 |
| 35 | 12.80 | 12.00 | .343 | .002 | 3.670 | .183 | .005 | .019 | .000 | .000 | .000 | .000 | .000 | .0000 | .0000 | .0000 | 29.74 | 2.41 |
| 35* | 12.80 | 12.00 | .353 | .002 | 5.650 | .335 | .010 | .271 | .112 | .022 | .020 | .182 | .005 | .0001 | .0039 | .0027 | 32.72 | 2.91 |
| 35** | 12.80 | 12.10 | .352 | .002 | 5.630 | .333 | .010 | .276 | .114 | .023 | .021 | .590 | .032 | .0013 | .0065 | .0015 | 33.24 | 2.85 |
| 40 | 13.70 | 13.90 | .340 | .002 | 2.610 | .076 | .002 | .004 | .000 | .000 | .000 | .000 | .000 | .0000 | .0000 | .0000 | 31.17 | 1.72 |
| 40* | 13.70 | 13.90 | .389 | .002 | 6.190 | .288 | .008 | .029 | .000 | .000 | .000 | .000 | .000 | .0000 | .0000 | .0000 | 35.84 | 3.72 |

Table 5 continued

| $m_*$ | H | He | C | N | O | Mg | Al | Si | S | Ar | Ca | Fe | Ni | Zn | Cr | Mn | $E_{tot}$ | $\Delta$ |
|---|---|---|---|---|---|---|---|---|---|---|---|---|---|---|---|---|---|---|
| $Z = 10^{-4} Z_\odot$ | | | | | | | | | | | | | | | | | | |
| 12 | 6.39 | 3.91 | .074 | .000 | .146 | .009 | .000 | .060 | .048 | .012 | .012 | .055 | .003 | .0004 | .0010 | .0002 | 10.78 | .56 |
| 13 | 6.82 | 4.31 | .109 | .000 | .181 | .008 | .000 | .052 | .028 | .006 | .005 | .090 | .004 | .0004 | .0016 | .0003 | 11.63 | .14 |
| 15 | 7.64 | 5.08 | .143 | .000 | .383 | .018 | .001 | .053 | .024 | .005 | .004 | .064 | .003 | .0003 | .0013 | .0002 | 13.45 | .24 |
| 18 | 8.80 | 6.12 | .219 | .000 | .766 | .058 | .002 | .098 | .038 | .007 | .006 | .162 | .008 | .0008 | .0019 | .0006 | 16.50 | 1.29 |
| 20 | 9.22 | 6.58 | .215 | .000 | 1.510 | .017 | .001 | .242 | .143 | .034 | .019 | .091 | .001 | .0000 | .0039 | .0006 | 18.14 | .34 |
| 22 | 9.95 | 7.42 | .273 | .000 | 1.730 | .033 | .001 | .280 | .167 | .036 | .023 | .124 | .003 | .0001 | .0040 | .0006 | 20.11 | .32 |
| 25 | 10.70 | 8.53 | .295 | .000 | 2.780 | .096 | .004 | .362 | .164 | .031 | .019 | .204 | .011 | .0004 | .0033 | .0008 | 23.29 | .38 |
| 30 | 12.00 | 10.10 | .333 | .000 | 4.070 | .254 | .008 | .098 | .003 | .000 | .000 | .000 | .000 | .0000 | .0000 | .0000 | 27.33 | 1.70 |
| 30* | 12.00 | 10.10 | .333 | .000 | 4.340 | .264 | .008 | .264 | .089 | .018 | .016 | .210 | .008 | .0003 | .0034 | .0006 | 28.11 | 1.62 |
| 35* | 13.20 | 11.80 | .358 | .000 | 5.550 | .330 | .009 | .097 | .003 | .000 | .000 | .000 | .000 | .0000 | .0000 | .0000 | 32.18 | 2.59 |
| 35** | 13.20 | 11.90 | .358 | .000 | 5.620 | .328 | .009 | .272 | .112 | .023 | .022 | .583 | .031 | .0014 | .0067 | .0014 | 33.30 | 2.50 |
| 35 | 13.20 | 11.70 | .164 | .000 | .026 | .000 | .000 | .000 | .000 | .000 | .000 | .000 | .000 | .0000 | .0000 | .0000 | 25.09 | .00 |
| 40 | 14.30 | 12.10 | .080 | .000 | .007 | .000 | .000 | .000 | .000 | .000 | .000 | .000 | .000 | .0000 | .0000 | .0000 | 26.49 | .01 |
| 40* | 14.30 | 13.60 | .420 | .000 | 6.200 | .284 | .007 | .036 | .001 | .000 | .000 | .000 | .000 | .0000 | .0000 | .0000 | 36.11 | 3.49 |
| 40** | 14.30 | 13.60 | .420 | .000 | 6.850 | .322 | .008 | .370 | .176 | .036 | .034 | .692 | .043 | .0019 | .0097 | .0023 | 38.12 | 3.29 |
| $Z = 0 Z_\odot$ | | | | | | | | | | | | | | | | | | |
| 12 | 6.39 | 4.08 | .043 | .000 | .067 | .004 | .000 | .025 | .009 | .002 | .001 | .081 | .009 | .0006 | .0007 | .0002 | 10.72 | .07 |
| 13 | 6.83 | 4.42 | .068 | .000 | .137 | .015 | .001 | .034 | .014 | .003 | .003 | .191 | .035 | .0029 | .0013 | .0006 | 11.79 | .29 |
| 15 | 7.69 | 4.90 | .145 | .000 | .400 | .034 | .001 | .069 | .026 | .005 | .004 | .161 | .014 | .0007 | .0011 | .0006 | 13.55 | .73 |
| 18 | 8.85 | 5.70 | .110 | .000 | .030 | .000 | .000 | .000 | .000 | .000 | .000 | .000 | .000 | .0000 | .0000 | .0000 | 14.69 | .00 |
| 20 | 9.56 | 6.32 | .090 | .000 | .009 | .000 | .000 | .000 | .000 | .000 | .000 | .000 | .000 | .0000 | .0000 | .0000 | 15.98 | .01 |
| 22 | 10.20 | 7.10 | .277 | .000 | 1.850 | .095 | .002 | .166 | .110 | .025 | .024 | .169 | .011 | .0007 | .0027 | .0008 | 20.60 | 2.75 |
| 25 | 11.40 | 7.33 | .083 | .000 | .010 | .000 | .000 | .000 | .000 | .000 | .000 | .000 | .000 | .0000 | .0000 | .0000 | 18.82 | -.02 |
| 30 | 12.80 | 9.11 | .109 | .004 | .027 | .000 | .000 | .000 | .000 | .000 | .000 | .000 | .000 | .0000 | .0000 | .0000 | 22.05 | .00 |
| 30* | 12.80 | 9.30 | .348 | .004 | 4.350 | .249 | .006 | .124 | .046 | .009 | .009 | .325 | .041 | .0031 | .0036 | .0008 | 28.67 | 3.67 |
| 35 | 14.20 | 8.17 | .000 | .000 | .000 | .000 | .000 | .000 | .000 | .000 | .000 | .000 | .000 | .0000 | .0000 | .0000 | 22.37 | .00 |
| 35* | 14.30 | 10.60 | .349 | .000 | 1.920 | .048 | .001 | .002 | .000 | .000 | .000 | .000 | .000 | .0000 | .0000 | .0000 | 27.60 | 1.38 |
| 35** | 14.30 | 10.70 | .404 | .000 | 5.580 | .239 | .005 | .211 | .090 | .018 | .018 | .578 | .032 | .0011 | .0060 | .0014 | 33.42 | 3.70 |
| 40 | 15.00 | 8.63 | .000 | .000 | .000 | .000 | .000 | .000 | .000 | .000 | .000 | .000 | .000 | .0000 | .0000 | .0000 | 23.63 | .00 |
| 40* | 15.20 | 12.00 | .285 | .000 | .578 | .000 | .000 | .000 | .000 | .000 | .000 | .000 | .000 | .0000 | .0000 | .0000 | 28.07 | .02 |
| 40** | 15.20 | 12.00 | .482 | .000 | 7.260 | .281 | .006 | .344 | .163 | .033 | .031 | .725 | .041 | .0025 | .0085 | .0018 | 38.17 | 4.17 |
| model A | | | | | | | | | | | | | | | | | | |
| 25 | 9.40 | 8.64 | .324 | .080 | 3.254 | .162 | .025 | .339 | .149 | .027 | .017 | .169 | .013 | .0015 | .0031 | .0031 | 23.07 | 2.01 |
| 25 | 10.30 | 8.60 | .267 | .009 | 2.900 | .032 | .003 | .399 | .223 | .054 | .022 | .207 | .011 | .0005 | .0036 | .0010 | 23.09 | .24 |
| 25 | 10.40 | 8.76 | .279 | .001 | 2.830 | .040 | .003 | .388 | .208 | .051 | .020 | .188 | .011 | .0004 | .0028 | .0008 | 23.24 | .24 |
| 25 | 10.70 | 8.53 | .295 | .000 | 2.780 | .096 | .004 | .362 | .164 | .031 | .019 | .204 | .011 | .0004 | .0033 | .0008 | 23.29 | .38 |
| 25 | 11.40 | 7.33 | .083 | .000 | .010 | .000 | .000 | .000 | .000 | .000 | .000 | .000 | .000 | .0000 | .0000 | .0000 | 18.82 | -.02 |
| SN Ia*** | | | | | | | | | | | | | | | | | | |
| | .00 | .00 | .048 | .000 | .143 | .009 | .001 | .153 | .086 | .016 | .012 | .744 | .141 | .0000 | .0064 | .0082 | | |

Notes: *model B, **model C, *** SN Ia data are from Nomoto *et al.* (1997) deflagration model W7

**Table 6.** stellar yields for various stellar masses ($m_*$) and three different initial metallicities Z = 1.0 $Z_\odot$ (= 0.02), Z = 0.2 $Z_\odot$ (= 0.004) and Z = 0.05 $Z_\odot$ (= 0.001) from van den Hoek & Groenewegen (1997)

| $m_*$ | H | H$^*$ | H$^{**}$ | He | He$^*$ | He$^{**}$ | C | C$^*$ | C$^{**}$ | N | N$^*$ | N$^{**}$ | O | O$^*$ | O$^{**}$ |
|---|---|---|---|---|---|---|---|---|---|---|---|---|---|---|---|
| 0.9 | -0.089 | -0.110 | -0.126 | 0.089 | 0.110 | 0.126 | -0.003 | -0.001 | 0.000 | 0.004 | 0.001 | 0.000 | 0.000 | 0.000 | 0.000 |
| 1.0 | -0.208 | -0.161 | -0.148 | 0.186 | 0.152 | 0.148 | -0.001 | 0.007 | 0.000 | 0.008 | 0.001 | 0.000 | 0.010 | 0.001 | 0.000 |
| 1.3 | -0.382 | -0.264 | -0.317 | 0.334 | 0.239 | 0.277 | 0.000 | 0.022 | 0.036 | 0.016 | 0.002 | 0.000 | 0.024 | 0.002 | 0.003 |
| 1.5 | -0.273 | -0.323 | -0.367 | 0.240 | 0.283 | 0.315 | 0.013 | 0.033 | 0.048 | 0.015 | 0.003 | 0.001 | 0.006 | 0.003 | 0.004 |
| 1.7 | -0.284 | -0.413 | -0.399 | 0.243 | 0.345 | 0.335 | 0.023 | 0.060 | 0.058 | 0.017 | 0.004 | 0.001 | 0.003 | 0.005 | 0.005 |
| 2.0 | -0.480 | -0.656 | -0.590 | 0.400 | 0.538 | 0.484 | 0.048 | 0.099 | 0.092 | 0.025 | 0.007 | 0.002 | 0.002 | 0.007 | 0.008 |
| 2.5 | -0.830 | -0.945 | -0.610 | 0.690 | 0.772 | 0.470 | 0.102 | 0.148 | 0.123 | 0.038 | 0.010 | 0.002 | -0.001 | 0.010 | 0.011 |
| 3.0 | -1.197 | -1.206 | -0.609 | 0.987 | 0.954 | 0.447 | 0.155 | 0.222 | 0.141 | 0.052 | 0.013 | 0.002 | 0.000 | 0.015 | 0.013 |
| 4.0 | -1.504 | -1.028 | -1.072 | 1.252 | 0.796 | 0.808 | 0.188 | 0.044 | 0.053 | 0.073 | 0.179 | 0.189 | -0.009 | 0.011 | 0.017 |
| 5.0 | -1.885 | -1.585 | -1.555 | 1.570 | 1.235 | 1.170 | -0.030 | 0.019 | 0.032 | 0.364 | 0.320 | 0.329 | -0.015 | 0.013 | 0.021 |
| 7.0 | -3.052 | -2.597 | -2.884 | 2.506 | 1.974 | 2.212 | -0.043 | 0.036 | 0.058 | 0.683 | 0.601 | 0.608 | -0.090 | -0.018 | 0.004 |
| 8.0 | -3.792 | -3.376 | -3.056 | 3.120 | 2.608 | 2.360 | -0.060 | 0.037 | 0.055 | 0.864 | 0.751 | 0.635 | -0.128 | -0.026 | 0.008 |

Notes: $^*$Z = 0.2 $Z_\odot$, $^{**}$Z = 0.05 $Z_\odot$